\documentclass[aps,prl,reprint,superscriptaddress]{revtex4-2}

\usepackage{blindtext}
\usepackage{amsmath}
\usepackage{amssymb}
\usepackage{graphicx}
\usepackage{appendix}
\usepackage{color}

\graphicspath{ {./Figs/} }

\newcommand{\vres}{{v_\mathrm{res}}}
\newcommand{\vthr}{{v_\mathrm{thr}}}
\newcommand{\vmin}{{v_\mathrm{min}}}

\newcommand{\taud}{{\tau_\mathrm{\delta}}}
\newcommand{\taum}{{\tau}}

\newcommand{\LFPadj}{\mathcal{L}^\dagger}
\newcommand{\psires}{{\vec{\psi}_\mathrm{res}}}

\newcommand{\CV}{c_v}
\newcommand{\ext}{\mathrm{ext}}

\newcommand{\EqRef}[1]{Eq.~\eqref{#1}}
\newcommand{\FigRef}[1]{Fig.~\ref{#1}}

\begin{document}

\title{Self-consistent stochastic dynamics for finite-size networks of spiking neurons}

\author{Gianni V. Vinci}
\affiliation{Natl. Center for Radiation Protection and Computational Physics, Istituto Superiore di Sanità, 00169 Roma, Italy}
\affiliation{PhD Program in Physics, “Tor Vergata” University of Rome, 00133 Roma, Italy}

\author{Roberto Benzi}
\affiliation{Dept. of Physics and INFN, “Tor Vergata” University of Rome, 00133 Roma, Italy}

\author{Maurizio Mattia}
\email{maurizio.mattia@iss.it}
\affiliation{Natl. Center for Radiation Protection and Computational Physics, Istituto Superiore di Sanità, 00169 Roma, Italy}

\date{December 29, 2021}

\begin{abstract} 
Despite the huge number of neurons composing a brain network, ongoing activity of local cell assemblies composing cortical columns is intrinsically stochastic. 
Fluctuations in their instantaneous rate of spike firing $\nu(t)$ scale with the size of the assembly and persist in isolated network, i.e., in absence of external source of noise. 
Although deterministic chaos due to the quenched disorder of the synaptic couplings likely underlies this seemingly stochastic dynamics, an effective theory for the network dynamics of a finite ensemble of spiking neurons is lacking.
Here, we fill this gap by extending the so-called population density approach including an activity- and size-dependent stochastic source in the Fokker-Planck equation for the membrane potential density.
The finite-size noise embedded in this stochastic partial derivative equation is analytically characterized leading to a self-consistent and non-perturbative description of $\nu(t)$ valid for a wide class of spiking neuron networks.
Its power spectra of $\nu(t)$ are found in excellent agreement with those from detailed simulations both in the linear regime and across a synchronization phase transition, when a size-dependent smearing of the critical dynamics emerges. 
\end{abstract}

\maketitle

\paragraph{Finite-size physical systems. ---} 

Complex systems in statistical physics usually deal with a huge number $N$ of interacting bodies like molecules leading to have wide applicability of effective mean-field theories.
However, the finite size (e.g., volume) of the system can have a significant impact in the emergent collective dynamics \cite{VanKampen2007}.
This can be due to the fact that finite systems have specific boundaries (finite surfaces and volumes) affecting the system behavior when phase transitions are approached \cite{Chen1997,Hwang2004}.
Besides, the presence of a finite number of elements can be incorporated as a stochastic field whose fluctuation size depend on $N$ \cite{Kawasaki1994,Dean1996}.
In this way, the continuum formalism valid in the thermodynamic limit may result to be effective in the description of small-scale phenomena across phase transitions \cite{DuranOlivencia2017}.
Biological and ecological systems can be even more challenging as 
they may incorporate peculiar boundary conditions and heterogeneity in space and time posing them continuously outside equilibrium \cite{Pascual1999,Gupta2014}.
This is the case of biological networks of neurons for which the probability current (see below) does not vanish even under stationary condition when it matches the frequency of spikes emitted per neuron, i.e. the firing rate $\nu$ \cite{Knight1972b,Abbott1993,Brunel1999,Fusi1999}.
In this Letter we will show that finite-size fluctuations can be effectively taken into account in this challenging system via a self-consistent definition of the noise to be embedded into the mean-field population dynamics.

\paragraph{Population density and mean-field approximation. ---}

In the thermodynamic limit ($N \to \infty$) networks of single-compartment spiking neurons have collective dynamics statistically described by the probability density $p(v,t)$ of realizations with membrane potential $v$ at time $t$ \cite{Abbott1993,Knight1996,Brunel1999,Fusi1999}, obeying the Fokker-Planck equation
\begin{equation}
\partial_{t}p = -\partial_v S_{p}
              = -\partial_{v}[(F+\mu) p] +\frac{1}{2} \partial^2_v (\sigma^2 p) \, .
\label{eq:FP}
\end{equation}
In this continuity equation the density $p$ changes according to the divergence of the probability current $S_p(v,t) = (F+\mu) p +\frac{1}{2} \partial_v (\sigma^2 p)$.
Here, the membrane potential $V(t)$ of a neuron follows the nonlinear Langevin equation
\begin{equation}
   dV = \left[F(V) + \mu\right]dt + \sigma dW \, ,
\label{eq:VDynDiff}
\end{equation}
where $F(V)$ is the drifting current determining the model-specific relaxation dynamics \cite{Gerstner2014}. 
In what follows, as workbench we adopt the `leaky' integrate-and-fire neuron with $F(V) = -V/\taum$ and decay time $\taum$.
The total synaptic and ionotropic input current is a Gaussian white noise $W(t)$ [$\langle W(t) W(t')\rangle=\delta(t-t')$] inhomogeneously modulated in time to have infinitesimal mean $\mu(V,t)$ and variance $\sigma^2(V,t)$.
Such diffusion approximation holds in the limit of large number $K$ of presynaptic contacts and small average synaptic efficacy $J$ \cite{Tuckwell1988}, well describing cortical networks \cite{DeFelipe2002,Markram2015}.
Following the Grigelionis central limit theorem \cite{Grigelionis1963}, this is due to having pooled together presynaptic spike trains with arbitrary inter-spike interval statistics provided that the $K$ input neurons are independent and fire at low rates $\nu$, eventually giving rise to a inhomogeneous Poisson process.
The independence hypothesis guaranteed by having small $J$, underlies also the mean-field approximation. 
In this framework all neurons are independent realizations of the same stochastic process (same $F$, $\mu$ and $\sigma$) \cite{Amit1997}.
Synaptic interactions are incorporated in the $\nu$-dependent moments of the current, which in current-based models like LIF neurons are 
\begin{equation}
\begin{array}{rcl}
	     \mu(\nu) & = & K J \nu(t) + \mu_\ext \\ 
   \sigma^2(\nu) & = & K J^2 \nu(t) + \sigma^2_\ext
\end{array} \, .
\label{eq:MuSigma}
\end{equation}
Here, the firing rate $\nu(t) = S_p(\vthr,t)$ is the flux of neurons crossing the emission threshold $\vthr$, and $\mu_\ext$ and $\sigma^2_\ext$ are the moments of the synaptic current due to the spikes incoming from neurons external to the network.

\paragraph{Why are finite-size fluctuations important? ---} 

Having a finite number $N$ of neurons emitting spike trains $S_i(t) = \sum_k{\delta(t-t_{i,k})}$, leads to a fluctuating rate $\nu_N(t) = \sum_{i=1}^N{S_i(t)}/N = \mathcal{N}(t)$ of action potentials fired per unit time and per neuron.
According to the aforementioned central limit, $\mathcal{N}(t)$ is an inhomogeneous Poission process with mean $N \, \nu(t)$, such that $\lim_{N\to\infty} \nu_N(t) = \nu(t)$.
Infinitesimal mean and variance of $\nu_N$ are $\nu$ and $\nu/N$, respectively, and for large enough $N$ ($\sim 100$) like those observed in cortical minicolumns \cite{Mountcastle1997}, holds the Gaussian approximation
\begin{equation}
   \nu_N(t) = \nu(t) + \sqrt{\frac{\nu(t)}{N}} W(t) \equiv \nu(t) + \eta(t) \, .
\label{eq:NuN}
\end{equation}
Here, the finite-size noise $\eta(t)$ results from a modulation of a white noise $W(t)$ independent from the one in \EqRef{eq:VDynDiff}. 
The correlation structure of $\eta$ will be self-consistently derived in the following.
Taking into account the fluctuating $\nu_N$ into \EqRef{eq:MuSigma}, the moments of the input current (i.e., the mean-field) are no longer deterministic \cite{Brunel1999}.
Fluctuations of the mean $\mu$ have relative size $\mathrm{Var}[\mu(\nu_N)]^{1/2}/\mathrm{E}[\mu(\nu_N)] = 1/\sqrt{N \nu}$ of about 10\% in cortical networks of interest where $\nu \sim 1$ Hz \cite{Watson2016}.
Remarkably, such variability is of the same order of the changes in $\mu$ associated with the coding of sensorial stimuli \cite{Poo2009} or other relevant information \cite{Rigotti2013}, which usually involves only a sparse set of tuned neurons with variations of few Hz in their firing rates. 
Thus, finite-size fluctuations may have a not negligible impact disturbing or nonlinearly amplifying the encoding dynamics of cortical networks.

\paragraph{How can finite-size fluctuations be incorporated? ---}

To understand the impact of such fluctuations, according to \cite{Brunel1999} we incorporate the stochastic moments $\mu_N \equiv \mu(\nu_N)$ and $\sigma_N \equiv \sigma(\nu_N)$ directly into \EqRef{eq:FP}.
The resulting stochastic Fokker-Planck equation describes now an infinite set of independent neurons all driven by the same fluctuating mean-field.
Differently from other stochastic Smoluchowski equations \cite{Kawasaki1994,Dean1996} and their coarse-grained versions \cite{Chavanis2008,Chavanis2015}, stochasticity here appears as an additional probability current with drift and diffusion coefficients differently affected by $\eta(t)$.
Not only, an additional source of noise must be incorporated as a fluctuating source of realizations in $v = \vres$ \cite{Mattia2002}. 
This is due to the fact that the finite flux of neurons crossing the threshold $\vthr$ reenters at the reset potential, eventually leading to 
\begin{equation}
\partial_{t}p = -\partial_{v}[(F+\mu_N) p]
                + \frac{1}{2} \partial^2_v (\sigma_N^2 p) 
					 + \delta(v - \vres) \, \eta(t) \, .
\label{eq:SFP}
\end{equation}
We remark that this stochastic Fokker-Planck (SFP) equation is nonlinear as both $\mu_N$ and $\sigma_N$ depend on the density $p$ via the firing rate $\nu$.

\paragraph{Self-consistent derivation of $\eta(t)$. ---}

To determine the statistical features of $\eta(t)$ we refer to the specific case of a set of $N$ uncoupled neurons ($J = 0$) driven by a stationary external input ($\dot{\mu}_\ext = \dot{\sigma}_\ext = 0$).
In this case neurons are renewal processes, and the probability density $\rho(t)$ of their inter-spike intervals (ISI) fully characterize the statistics of the spike trains they emit \cite{Cox1977}.
Pooling together these spike trains gives $\nu_N(t)$ (see above) such that its power spectral density is \cite{Lindner2006,Gerstner2014} 
\begin{equation}
   P_{\nu}^{(\mathrm{RT})}(\omega) = |\nu_N(\omega)|^2 = \frac{\nu_0}{N} \, \mathrm{Re}\left[\frac{1+\rho(\omega)}{1-\rho(\omega)} \right] \, .
\label{eq:PSDofNuFromRT}
\end{equation}
Here $f(\omega) = \int_{-\infty}^\infty{f(t) e^{-i\omega t} dt}$ is the Fourier transform of any function $f(t)$, and $\nu_0 = 1/\langle T \rangle$ is the mean firing rate, i.e., the inverse of the mean ISI.

This exact result must be also obtained from \EqRef{eq:SFP}.
To carry out the power spectral density $|\nu_N(\omega)|^2$ in this case we resort to the spectral expansion approach introduced in \cite{Mattia2002} giving
\begin{equation}
   P_{\nu}^{(\mathrm{SE})}(\omega) =  \, \left| 1 + \vec{f} \cdot (i \omega \mathbf{I} - \mathbf{\Lambda})^{-1} \psires \right| |\eta(\omega)|^2
\label{eq:PSDofNuFromSE}
\end{equation}
If the finite-size noise is assumed to be white, $|\eta(\omega)|^2 = \nu_0/N$ leading to an overestimate of the power $P_\nu(\omega)$ at relatively low-$\omega$ at least under `suprathreshold' regime, i.e., when neurons emit spikes even if $\sigma = 0$ \cite{Mattia2002}.

In \EqRef{eq:PSDofNuFromSE}, we make use of the eigenfunctions $\phi_n(v)$ of the non-Hermitian Fokker-Planck operator $\mathcal{L}$ defined from \EqRef{eq:FP} as $\partial_t p \equiv \mathcal{L} \, p$. 
This operator has an infinite spectrum of discrete eigenvalues $\lambda_n$ ($n \in \mathbb{Z}$) such that $\mathcal{L} \, \phi_n = \lambda_n \, \phi_n$ \cite{Abbott1993,Knight1996,Mattia2002}.
The flux $f_n = 1/2 \partial_v (\sigma \phi_n)|_{v=\vthr}$ of nonstationary ($n \neq 0$) eigenfunctions are the infinite elements of $\vec{f}$, while the matrix $\mathbf{\Lambda}$ is diagonal with $\{\mathbf{\Lambda}\}_{nn} = \lambda_n$.
The elements of $\psires$ are instead the eigenfunction $\psi_n(v)$ of the adjoint operator $\LFPadj$ with same $\lambda_n$, computed in $v = \vres$ \cite{Knight1996,Mattia2002}. 
Note that all these coefficients are state-depend being functions of the current moments $\mu$ and $\sigma$.

\begin{figure}[ht!]
\includegraphics[width=\columnwidth]{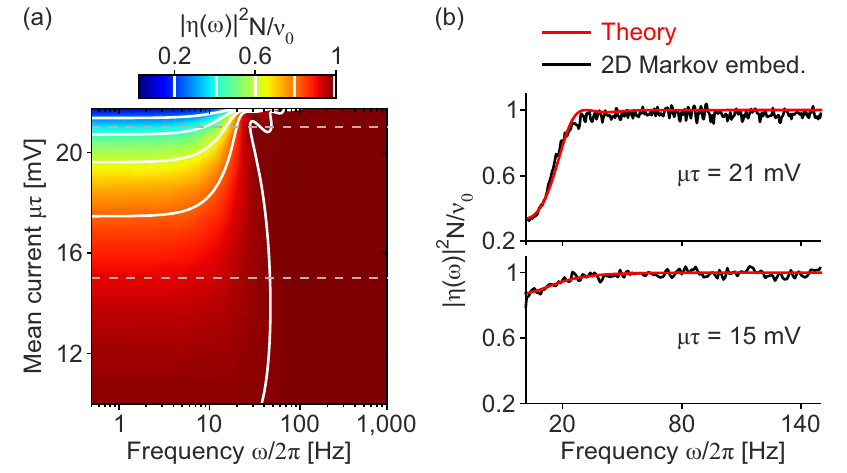}
\caption{Normalized power spectra of the finite-size noise $\eta$ for $N$ uncoupled LIF neurons with firing rate $\nu_0$. 
(a) $|\eta(\omega)|^2 N/\nu_0$ as a function of the mean synaptic current $\mu$. 
$\sigma$ is chosen to keep the mean firing rate unchanged at $\nu_0 = 20$~Hz. 
(b) In noise-dominate regime ($\mu\taum < \vthr = 20$~mV) finite-size noise is essentially white (bottom). 
In drift-dominate regime, power is low at low-$\omega$ as ISIs are more regular ($\CV < 1$). 
In both cases, a two-dimensional Markovian embedding (black) faithfully reproduce theoretical spectra (red) from \EqRef{eq:PSDofEta}.}
\label{fig:FSnoise}
\end{figure}

For any IF neuron model, the series in \EqRef{eq:PSDofNuFromSE} can be summed as a function of $\rho(\omega)$ \cite{Vinci2021a}: 
\begin{equation}
	\vec{f} \cdot (i \omega \mathbf{I} - \mathbf{\Lambda})^{-1} \psires = 
	\frac{\rho(\omega)}{1-\rho(\omega)} - \frac{\nu_0}{i \omega} \, .
\label{eq:HofS}
\end{equation}
Using it in \EqRef{eq:PSDofNuFromSE} and requiring the equivalence $P_{\nu}^{(\mathrm{SE})}(\omega) = P_{\nu}^{(\mathrm{RT})}(\omega)$, we obtain with \EqRef{eq:PSDofNuFromRT} a self-consistent expression for the power spectrum of $\eta$:
\begin{equation}
   |\eta(\omega)|^2 = \frac{\nu_0}{N} \left[ 1 - \left| \frac{(i\omega +\nu_0)\rho(\omega) -\nu_0 }{\nu_0 \rho(\omega) +i\omega -\nu_0} \right|^2 \right] \, .
\label{eq:PSDofEta}
\end{equation}
In the limit $\omega \to \infty$, $\rho(\omega) \to 0$ and the finite-size noise is white-like with variance $\nu_0/N$.
For  $\omega \to 0$, the l.h.s. of \EqRef{eq:HofS} reduces to 
$-\vec{f} \cdot \mathbf{\Lambda}^{-1} \psires = (\CV^2 - 1)/2$ \cite{Vinci2021a}, function of 
the coefficient of variation $\CV$ of the ISIs, leading to
\begin{equation}
|\eta(0)|^2 = \frac{\nu_0}{N}\frac{4 \CV^2}{(1+ \CV^2)^2} \, .
\end{equation}
These limits suggest a sigmoid-like power spectra of the finite-size noise which is confirmed in LIF neurons [\FigRef{fig:FSnoise}(a)]. 
As the firing regimes moves from noise- to drift-dominated (i.e., from sub- to suprathreshold) regimes by increasing the mean current $\mu$, the sigmoidal shape becomes increasingly more apparent.
Indeed, ISIs are more and more regular leading to a decrease of their $\CV$, and hence to a lower power $|\eta(0)|^2$.

\paragraph{Markovian embedding of $\eta(t)$. ---}

Having derived in \EqRef{eq:PSDofEta} the correlation structure of the finite-size noise for a set of uncoupled neurons under stationary condition a question arises: can we generalize this result to networks of synaptically coupled neurons and far from equilibrium?

To answer this question we remark that the spectra from \EqRef{eq:PSDofEta} shown in \FigRef{fig:FSnoise} are constant (white noise) with a power reduction at low-$\omega$ possibly resulting by subtracting a Lorentzian-shaped function.
Following \cite{Vellmer2019}, such kind of spectra are very well approximated by two-dimensional Ornstein-Uhlenbeck processes $\vec{u}(t)$ driven by and interfering with the same white noise $W(t)$:
\begin{equation}
\begin{array}{rcl}
   d\vec{u} & = & \mathbf{A} \, \vec{u} \, dt + \mathbf{B} \, dW \\
       \eta & = & \displaystyle \vec{1} \cdot \vec{u} + \sqrt{\frac{\nu_0}{N}} \, W
\end{array} \, .
\label{eq:MarkovEmbedding}
\end{equation}
In \FigRef{fig:FSnoise}(b) we show that this Markovian embedding faithfully reproduces the correlation structure of $\eta(t)$ in the simple case of $\mathbf{B}$ with only one non-zero element, $B_{11}=\sqrt{\nu_0/N}$, and of a three-parameter $\mathbf{A}$ defined as $A_{11} = A_{22} = a$, $A_{12} = a_\uparrow$ and $A_{21} = a_\downarrow$.
Instead of fitting $\{a,a_\uparrow,a_\downarrow\}$, we carry them out analytically as in \cite{Vellmer2019} by matching the power of $\eta$ from \EqRef{eq:MarkovEmbedding} with the exact one in \EqRef{eq:PSDofEta} at the frequencies: $\omega = \{0,\pi\nu_0,2\pi\nu_0\}$.
The parameters change according to $\mu$ and $\sigma$ leading to a state-dependent $\mathbf{A}(\mu,\sigma)$.

This dynamical description of $\eta(t)$ in principle allows to overcome the renewal hypothesis, as the memory embedded in the network activity is reintroduced via the dependence on $\nu(t)$ of the current moments in \EqRef{eq:MuSigma}.
We then conjecture that taken together Eqs.~\eqref{eq:SFP} and \eqref{eq:MarkovEmbedding} provide a complete statistical description of the out-of-equilibrium dynamics of a finite-size network of spiking neurons.

\paragraph{Interacting neurons and linear-response theory. ---}

\begin{figure}[ht!]
\includegraphics[width=\columnwidth]{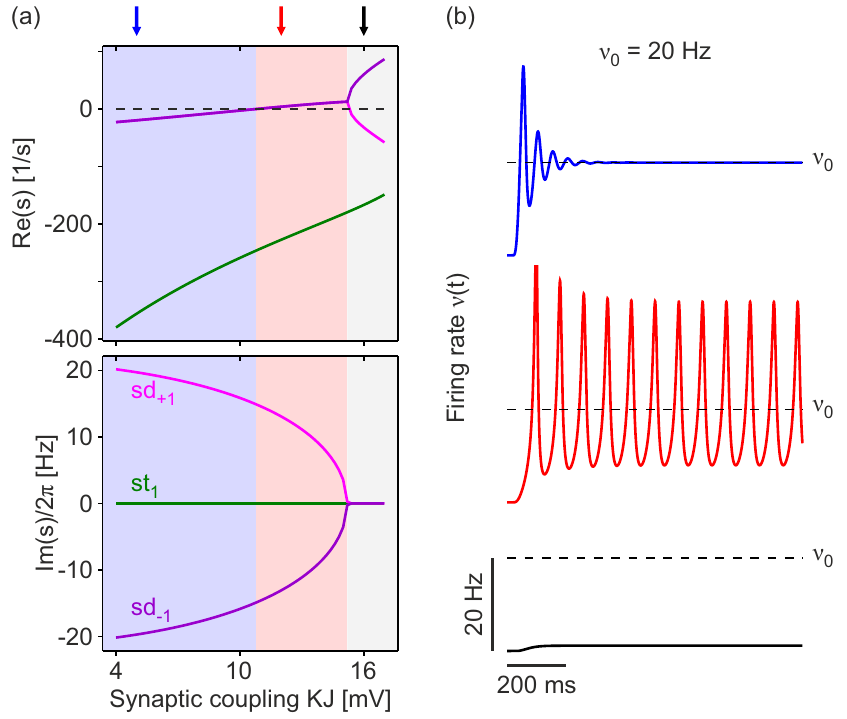}
\caption{Bifurcation analysis of a network of LIF excitatory neurons ($N \to \infty$) at varying synaptic efficacy $J$. 
(a) Real and imaginary parts of the poles $\mathrm{sd}_{\pm 1}$ and $\mathrm{st}_1$ of $\nu(s)$ solving \EqRef{eq:PolesOfNuLT}. 
Weakly coupled networks (small $KJ$, blue range) have a stable focus in $\nu_0$. 
Beyond $KJ \simeq 11$~mV (red) limit cycles arise \textit{via} a supercritical Hopf bifurcation, eventually crossing a saddle-node bifurcation (grey).
(b) Numerical integration of the Fokker-Planck equation with $p(v,0)=\delta(v-\vres)$ for $KJ$ pointed out in (a) (top arrows).
The persistence of the fixed-point $\nu_0 = 20$~Hz results from keeping constant the moments $\mu \, \taum = 21$~mV and $\sigma \, \taum^{1/2} = 2.665$~mV, and changing $\mu_\ext$ and $\sigma_\ext$ according to $J$. 
Other parameters: $K = 10^3$, $\taum = 20$~ms, $\vthr = 20$~mV, $\delta_\mathrm{min}=2$~ms and $\taud = 1$~ms.}
\label{fig:BifurcDiagr}
\end{figure}

We test this conjecture using as test bench a network of LIF neurons with a fixed-point at $\nu_0 = 20$~Hz in the thermodynamic limit ($N \to \infty$).
We balance the increase in the excitatory synaptic efficacy $J$ varying the moments $\mu_\ext$ and $\sigma_\ext$ of the external current to keep $\nu_0$ unchanged.
An exponential distribution of axonal delays $\delta > \delta_\mathrm{min}$ in delivering emitted spikes is taken into account and having the same impact as non-instantaneous synaptic transmission \cite{Mattia2019}.
The incoming rate of spikes $\tilde{\nu}(t)$ has dynamics $\dot{\tilde{\nu}} = (\nu-\tilde{\nu})/\taud$, and replaces $\nu(t)$ in \EqRef{eq:MuSigma}.
The stability analysis of the equilibrium point $\nu_0$ of the network can be effectively carried out resorting to the spectral expansion of $p(v,t)$ mentioned above.
Indeed, the linear dynamics of the perturbations $\nu(t)-\nu_0$ is fully captured by the time course of the projections on the two slowest eigenmodes \cite{Mattia2021}.
Stability are then determined by the poles $s$ of the Laplace transform $\nu(s)=\int_0^\infty{e^{-s t} \nu(t) dt}$ of the firing rate solving the equation \cite{Mattia2002}
\begin{equation}
   1 - g(s) \left[ \Phi' +\frac{s}{\taum}\left(\frac{c_1}{s-\lambda_1} + 
                                               \frac{c_2}{s-\lambda_2} \right) 
            \right] = 0 \, .
\label{eq:PolesOfNuLT}
\end{equation}
Here $g(s) = \frac{e^{-(s-\delta_\mathrm{min})/\tau_\delta}}{\tau_\delta}$ is the Laplace transform of the delay distribution, $\Phi' = \partial_\nu \Phi$ is the slope of the current-to-rate gain function $\Phi(\mu,\sigma) = S_{\phi_0}(\vthr)$ computed in $\nu_0$, and $c_n = \int_\vmin^\vthr{\phi_0 \partial_\nu \psi_n dv}$ with $n \in \{1,2\}$ are coupling coefficients proportional to $J$ \cite{Mattia2002,Mattia2021}. 
Real and imaginary parts of the three poles solving \EqRef{eq:PolesOfNuLT} are shown in \FigRef{fig:BifurcDiagr}(a) as a function of $KJ$ .
Three phases are singled out starting from a stable focus in $\nu_0$ for weak couplings which with increasing $KJ$ destabilizes via a supercritical Hopf bifurcation giving rise to a stable limit cycle. 
Large enough $KJ$ eventually lead to a saddle-node bifurcation where $\nu(t)$ is asymptotically attracted to low firing rates.
The numerical integrations \cite{Augustin2017} of the Fokker-Planck \EqRef{eq:FP} for three representative networks are shown in \FigRef{fig:BifurcDiagr}(b) confirming the reliability of the approximated bifurcation analysis.

\paragraph{Linear response regime for finite networks. ---}

\begin{figure}[ht!]
\includegraphics[width=\columnwidth]{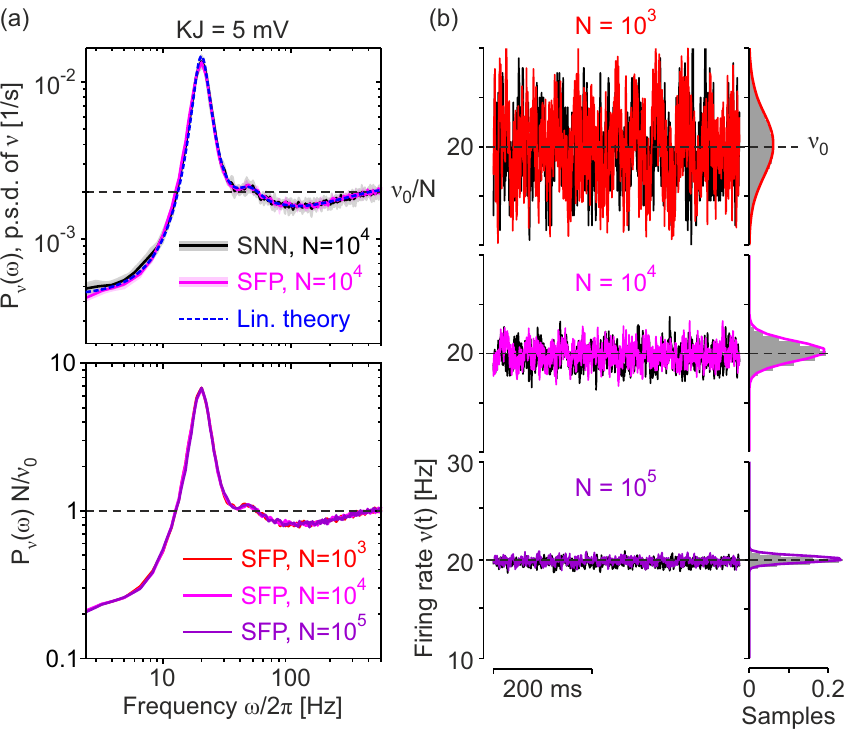}
\caption{Power spectra $P_{\nu}(\omega)$ (a) and firing rates $\nu_N(t)$ (b) of weakly coupled networks ($KJ = 5$~mV) with varying size $N$.
(a) Top, $P_{\nu}(\omega)$ from the numerical integration of the stochastic Fokker-Planck equation (SFP, pink), the equivalent spiking neuron network (SNN) simulation (black) and linear perturbation theory (dashed blue) ($N = 10^4$).
Dashed black, white noise with variance $\nu_0/$.
Bottom, normalized spectra of $\nu$ from SFP of networks with different $N$ ($= 10^3, 10^4, 10^5$). 
Theoretical power spectrum (dotted blue) is from \EqRef{eq:LinearPSDofNu}. 
(b) $\nu(t)$ from SNN simulations (black) and SFP varying $N$ ($= 10^3, 10^4, 10^5$) (left). 
Right, related histograms of $\nu(t)$.
Other parameters as in \FigRef{fig:BifurcDiagr}.}
\label{fig:WeaklyCoupled}
\end{figure}

In weakly coupled networks with relatively small $KJ$, finite-size noise induces stochastic perturbations of $\nu_N(t)$ around an equilibrium point amenable to linear response theory.
This is the case of our example network with $KJ = 5$~mV for which the power spectrum of $\nu_N$ can be carried out as a linear transformation of the finite-size noise \cite{Mattia2002,Mattia2004}: 
\begin{equation}
   P_{\nu}(\omega) = |\nu_N(\omega)|^2 = |H_\nu(\omega)|^2 |\eta(\omega)|^2 \, .
\label{eq:LinearPSDofNu}
\end{equation}
Here $H_\nu(\omega)$ is the Fourier transfer function giving the linear response to sinusoidal modulations of the input firing rate which is analytically known for LIF neurons \cite{Brunel1999,Vinci2021a}.
In \FigRef{fig:WeaklyCoupled}(a)-top we show a remarkable agreement between such theoretical expression, and $P_{\nu}(\omega)$ estimated from both the simulations of spiking neuron networks (SNN, \cite{NEST2020}) and the numerical integration of the SFP \EqRef{eq:SFP}. 
To computed the SFP integration we extended a standard deterministic approach \cite{Augustin2017} by incorporating the Markovian embedding \eqref{eq:MarkovEmbedding} of $\eta(t)$.
SFP integration for different network sizes $N$  confirms in \FigRef{fig:WeaklyCoupled}(a)-bottom what expected from the linear theory, that is, the spectrum shape does not change with $N$ once normalized by the variance $\nu_0/N$.
The equivalence in these case between SNN simulation and SFP integration is also apparent in the direct comparison of $\nu_N(t)$ time series.
Indeed, in \FigRef{fig:WeaklyCoupled}(b) no differences emerge and both time series display a variance scaling as $1/N$.

\paragraph{Beyond linear response theory. ---}

\begin{figure*}[ht!]
\includegraphics[width=\textwidth]{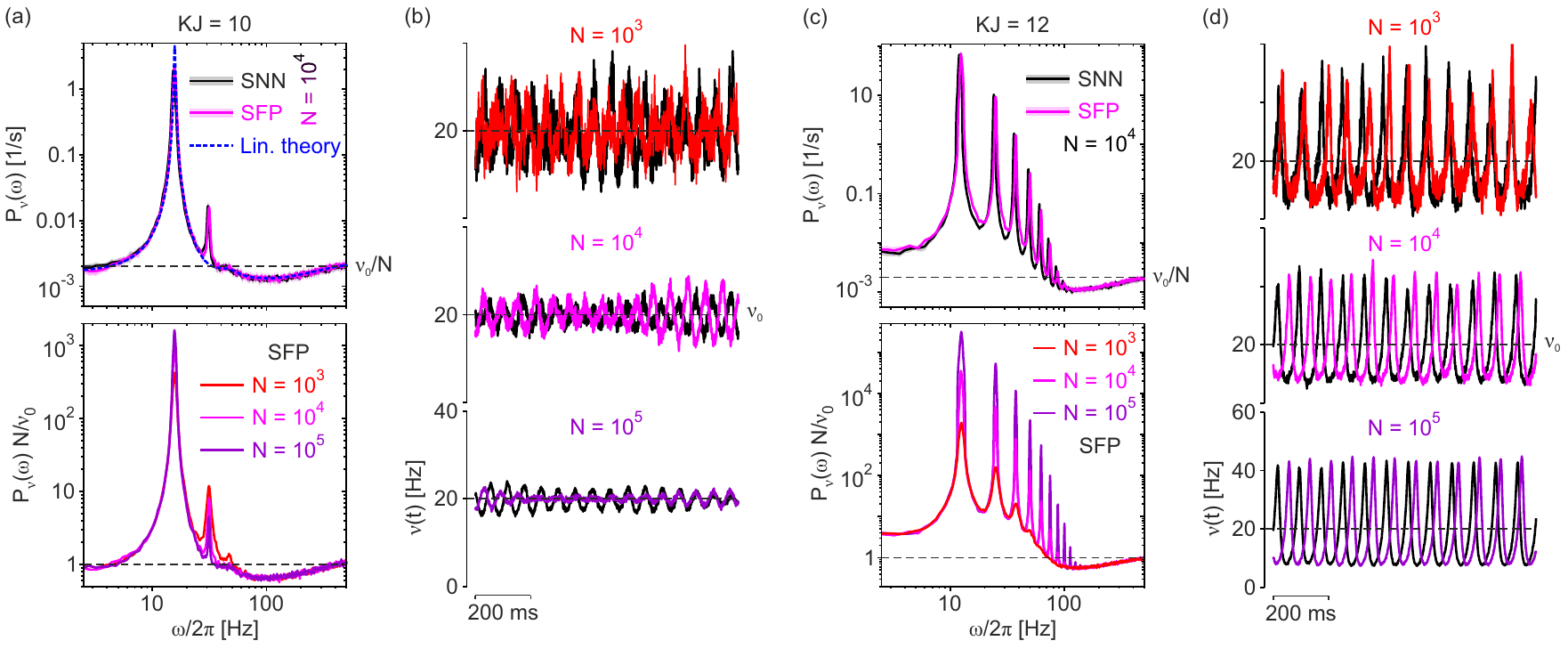}
\caption{
Power spectra $P_{\nu}(\omega)$ (a,c) and firing rates $\nu_N(t)$ (b,d) of strongly coupled (nonlinear) networks with varying size $N$. 
(a, b) $KJ = 10$~mV, mean-field dynamics predicts a stable focus at $\nu_0 = 20$~Hz. 
(c, d) $KJ = 12$~mV, networks beyond a supercritical Hopf bifurcation with a stable limit cycle.
See Figs.~\ref{fig:BifurcDiagr} and \ref{fig:WeaklyCoupled} for additional details.}
\label{fig:StronglyCoupled}
\end{figure*}

By further increasing synaptic coupling $KJ$, a supercritical Hopf bifurcation is approached (\FigRef{fig:BifurcDiagr}) and finite-size networks have $P_\nu(\omega)$ no longer fully described by linear theory [i.e., \EqRef{eq:LinearPSDofNu}].
The mismatch is apparent in \FigRef{fig:StronglyCoupled}(a)-top as an unpredicted  second-harmonic peak at $\omega/2\pi \simeq 2 \nu_0$ arises both in SNN simulations and in the SFP integration.
Despite such footprint of nonlinear dynamics, SNN and SFP keep displaying a remarkable overlap not only in $P_\nu$ but also in the stochastic dynamics of $\nu_N(t)$ shown in \FigRef{fig:StronglyCoupled}(b) where coherent oscillations become more and more apparent as $N$ decreases.
The ongoing finite-size fluctuations of $\nu$ in this case continuously stimulate the oscillating relaxation of the stable-focus which is relatively slow due to the nearby critical point.
The $N$-dependent coherence of the fluctuation-driven oscillation is even more apparent in the increase of power of the second-harmonic peak in \FigRef{fig:StronglyCoupled}(a)-bottom as the size of the network is reduced.

The same remarkable match between SFP and SNN is shown in \FigRef{fig:StronglyCoupled}(c)-top beyond the Hopf bifurcation ($KJ = 12$~mV) when a stable limit cycle is expected at $N \to \infty$.
Global oscillations in this case are strongly nonlinear giving rise to several high-order harmonic peaks and finite-size noise contribute to dephase them limiting their coherence in time [\FigRef{fig:StronglyCoupled}(d)].
According to this, the resonant peaks in $P_\nu(\omega)$ have power lowering with decreasing $N$ [\FigRef{fig:StronglyCoupled}(c)-bottom].

\paragraph{Conclusions. ---}

The agreement found between SFP and SNN in a synchronization phase transition is an encouraging evidence that the stochastic population dynamics we derived in Eqs.~\eqref{eq:SFP}, \eqref{eq:MarkovEmbedding} and \eqref{eq:NuN} has the potential to faithfully describe out-of-equilibrium networks composed of a finite number $N$ of spiking neurons.
Indeed, it is important to stress that the theoretical framework we introduce is not perturbative and takes into account the point-like (spike-based) nature of the cell-to-cell interactions.
Furthermore, the correlation structure of $\eta(t)$ here results from the spiking statistics of isolated neurons, implying that finite-size noise is not $J$- dependent.
Not only, its Markovian embedding allows in principle to derive single-neuron ISI statistics also under nonstationary conditions, i.e., when renewal hypothesis is broken.

Alternative approaches dealing with finite-size networks of spiking neurons are those relying on the refractory density method (RDM) \cite{Schwalger2019}.
Neurons in this case are quasi-stationary renewal processes and the population dynamics is fully described by the probability density of a single-cell to be into a refractory state \cite{Gerstner1995,Gerstner2000}.
In the RDM, the integro-differential equation governing the population activity at the mesoscopic scale can incorporate fluctuations to describe a finite number of neurons \cite{Schwalger2017,Schmutz2021}.
However, the renewal hypothesis underlying RDM may in principle limit its applicability by making our theoretical framework preferable in dealing with out-of-equilibrium conditions. 

In conclusion, we remark that the theoretical framework we developed applies to a wide class of spiking neuron models, and the same approach can be extended to other physical systems with a finite size.
Indeed, the inclusion of fluctuations in population density theories has the potential to further advance our understanding of noise-driven out-of-equilibrium dynamics \cite{Das2012,DuranOlivencia2017}.

Besides, numerical integrations of SFP in the examples shown computationally outperform detailed simulations of SNNs. 
In principle, this paves the way to implement large-scale heterogeneous networks of neuronal population making affordable detailed \textit{in silico} experiments on macroscopic brain regions \cite{SanzLeon2013}.

\paragraph{Acknowledgments. ---}

Work in part funded by EU H2020 Research and Innovation Programme, Grant 945539 (HBP SGA3) to MM.

\bibliographystyle{plain}
\bibliography{VinciBenziMattia}

\end{document}